\documentclass[pra,10pt, twocolumn]{revtex4-1}
\usepackage{makecell,multirow,diagbox}
\usepackage{amssymb,amsmath,graphicx,color}
\usepackage{times}

\begin{document}

\title{Single-step multipartite entangled  states generation from coupled circuit cavities}
\author{Xiao-Tao Mo}
\author{Zheng-Yuan Xue} \email{Email: zyxue83@163.com}
\affiliation{Guangdong Provincial Key Laboratory of Quantum Engineering
and Quantum Materials, and School of Physics\\ and Telecommunication Engineering,
South China Normal University, Guangzhou 510006, China}

\date{\today}

\begin{abstract}
Green-Horne-Zeilinger states are a typical type of multipartite entangled states, which plays a central role in quantum information processing. For the generation of multipartite entangled states, the single-step method is more preferable as the needed time will not increase with the  increasing of the qubit number. However, this scenario has a strict requirement that all the two-qubit interaction strengths should be the same, or the generated state will be of low quality.  Here, we propose a scheme for generating multipartite entangled  states of superconducting qubits, from a coupled circuit cavities scenario, where we rigorously achieve the requirement via  adding an extra \emph{z}-direction ac classical field for each qubit, leading the individual qubit-cavity coupling strength to be tunable in a wide range, and thus can be tuned to the same value. Meanwhile, in order to obtain our wanted multi-qubits  interaction, \emph{x}-direction ac classical field for each qubit is also introduced. By selecting the appropriate parameters,  we numerically shown that  high-fidelity multi-qubit GHZ states can be generated. In addition, we also show that the coupled cavities scenario is better than a single cavity case. Therefore, our proposal represents a promising alternative for multipartite entangled states generation.
\end{abstract}

\maketitle

\section{INTRODUCTION}

Entanglement is one of the most counterintuitive consequences of quantum physics, and nowadays it plays a central role in quantum computation and quantum communication \cite{qc}. Multipartite entangled states are entangled states of many qubits which are indispensable resource for research in large scale quantum computation \cite{t1}, multipartite quantum communication \cite{commun}, quantum simulation \cite{qs} and quantum-to-classical transition \cite{qf}. Therefore, generating entangled states of an increasing number of qubits is an important benchmark for modern quantum technology \cite{qt}.

Green-Horne-Zeilinger (GHZ) states \cite{ghz} are a typical type of maximally entangled states, and the generation of which have been paid much attention recently \cite{e0,e1,sbs,e2,e3,e41,e4,e42,e5,e6,e7,e8,e9,qi}. Conventional way of GHZ states is generated in a step by step way, based on high-fidelity quantum gates. In this way, the number of entangled qubits is only increased by one at a time, and thus the needed generation time will increase with the increasing of the number of the involved qubits. In addition, this method will also result in accumulation of individual gate operation errors when the number of entangled qubits increases. Alternatively, GHZ state can be generated in a single step \cite{ss1,ss2,ss3,ss4,ss5,ss6,ss7,ss8,ss9,ss10,ss11,ss12,ss13,ss14}, via the deliberately designed collective interaction, irrespective of the number of entangled qubits.

Meanwhile, superconducting transmon qubits, a kind of superconducting Josephson-junction qubit, are one of the promising solid-state processor for quantum state manipulation \cite{sq1,sq2,sq3}. Recently, the setup of multipartite superconducting qubits connecting to a common bus resonator has been used for single-step entangled state generation. But, with the increasing of the number of qubits, the probability to generate a GHZ state decrease dramatically. This is because that the single-step method has a strict requirement that all the two-qubit interaction strengths should be the same. However, for each qubit, its coupling strength with the bus resonator is different so that the qubit-qubit interaction does not evolve synchronously.

Here, we present a scheme to solve the above-mentioned difficulty in a two coupled cavities scenario, where each superconducting qubit is biased by a \emph{z}-direction magnetic flux in order to tune the qubit-resonator coupling. Specifically, this modulation can effectively make the qubit-resonator couplings to be tunable via the amplitudes of the external driving fields, so that the same coupling strength requirement for many qubits can be met. Meanwhile, in order to obtain our wanted multi-qubits  interaction, different from that of in Refs. \cite{ss1,ss2,ss12,ss13}, \emph{x}-direction ac classical field for each qubit is also introduced. By selecting the appropriate parameters,  we analytically proved that multi-qubit GHZ states can be generated and we also numerically simulated the obtained high fidelity. In addition, for the target GHZ generation task, we also numerically show that the coupled cavities scenario is better than a single cavity case. Therefore, our proposal represents a promising alternative for multipartite entangled GHZ states generation with superconducting qubits.

\section{The theoretical scheme}
The proposed setup for Generating GHZ state is illustrated in Fig. 1, which consists of a two coupled cavities in the circuit QED \cite{sq2} scenario. Setting $\hbar=1$ hereafter, the Hamiltonian of the two coupled cavities is
\begin{eqnarray}\label{1}
H_{c}&=&\omega_ra^+a+\omega_rb^+b+J(ab^++a^+b)  \nonumber\\
     &=&\omega_+P_+^+P_++\omega_-P_-^+P_-,
\end{eqnarray}
where   $a^{\dagger}$ ($b^{\dagger}$) and $a$ ($b$) are the creation and annihilation operators for the cavity A (B), respectively; $J$ is the coupling strength between the two cavity modes. The two localized normal modes of this coupled system are $P_{\pm}=(a\pm b)/\sqrt{2}$, and the frequencies of them are $\omega_{\pm}=\omega_r\pm J$, with $\omega_r$ being chose to be the same  for simplicity. Otherwise, the two modes and their frequency will be slightly modified.

\begin{figure}[tbp]
  \centering
  \includegraphics[width=\linewidth]{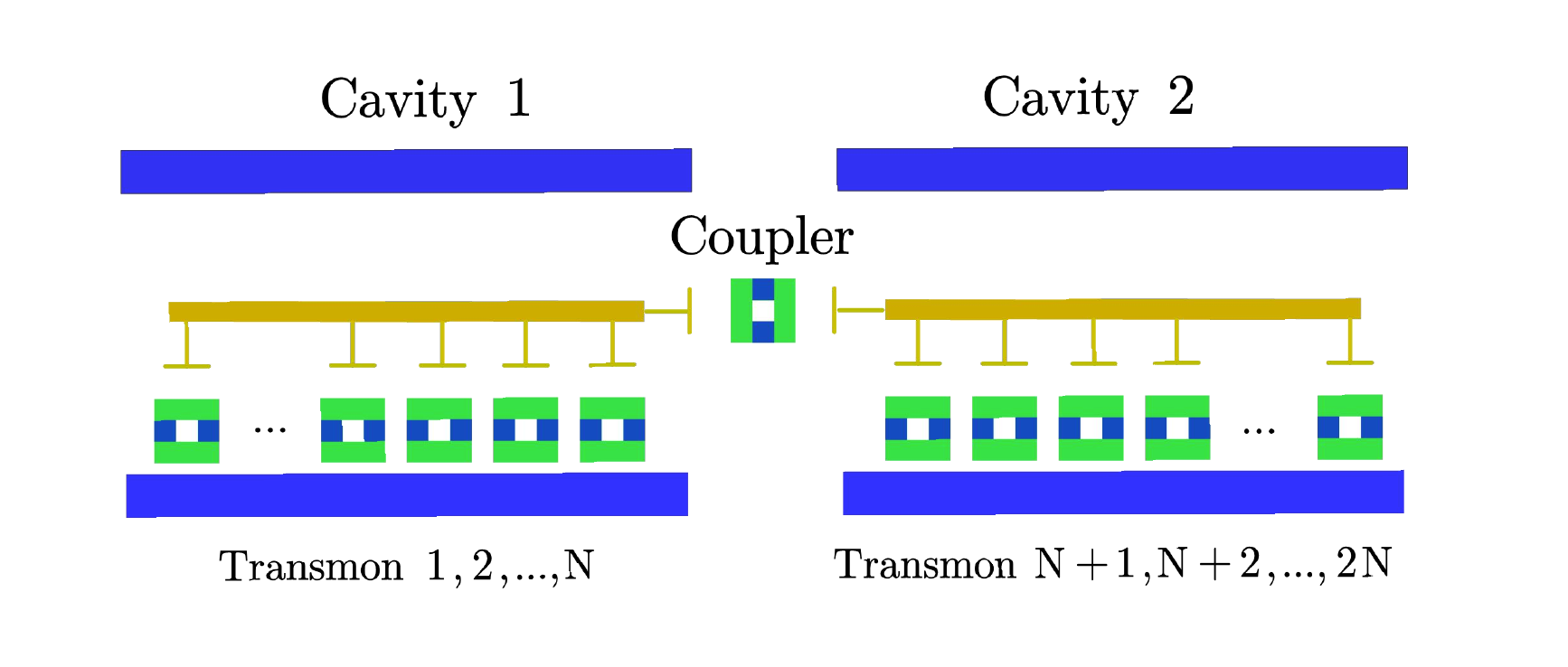}
  \caption{ Schematic diagram of our proposal. The system consists of two cavities coupled by a central coupler, where \emph{N} transmons qubits (the yellow stripe) are placed in and coupled to each  cavity.}\label{ene}
\end{figure}

In each cavity, there are N transmon qubits, and the free Hamiltonian of them are
\begin{eqnarray}\label{2}
H_q=\sum_{j=1}^{2N}\frac{\omega_{q_j}}{2}\sigma_j^z,
\end{eqnarray}
where the static qubit frequencies are $\omega_{q,j}$. The 2\emph{N} qubits are simultaneously coupled to their corresponding cavities, and the coupling Hamiltonian can be generally written as
\begin{eqnarray}\label{3}
H_{int}=\sum_{j=1}^{N}\frac{g_j}{2} a^+\sigma_j^- +\sum_{j=N+1}^{2N}\frac{g_j}{2} b^+\sigma_j^- +H.c.,
\end{eqnarray}
where $\sigma_j^z=|1\rangle_j\langle1|-|0\rangle_j\langle0|$,  $\sigma_j^+=|1\rangle_j\langle0|$, $\sigma_j^-=|0\rangle_j\langle1|$,
 with $|0\rangle_j$ and $|1\rangle_j$ being the ground and excited states of $j$th qubit, $g_j$ is the $j$th qubit-cavity coupling strength.

Meanwhile, all the qubits are simultaneously driven by the classical field along \emph{x} and \emph{z} directions as
\begin{eqnarray}\label{4}
H_{dz} &=& \sum_{j=1}^{2N} \frac{A_j\sin(\omega_j t+\varphi)}{2}\sigma_j^z, \notag\\
H_{dx}&=&\sum_{j=1}^{2N} \frac{\Omega_j}{2}\left[e^{-i\omega_{d_j} t} \sigma_j^+ +e^{i\omega_{d_j} t}\sigma_j^-\right],
\end{eqnarray}
where $\Omega_j$ is the Rabi frequency of the classical field along \emph{x} direction and $A_j$ is the amplitude of the classical field which can drive jth qubit along \emph{z} direction.

In the interaction picture with respect to $H_0=H_c+H_{dz}+H_q$,
the interaction Hamiltonian $H_w=\exp{(iT\int H_0 dt)} (H_{int}+H_{dx}) \exp{(-i T\int H_0 dt)}$, with $T$ being the time-ordering operator,  will be
\begin{widetext}
\begin{eqnarray}\label{hw}
H_w&=&\sum_{j=1}^{N}\frac{g_j}{2\sqrt{2}}(P_+^+ e^{i\omega_+ t} +P_-^+e^{i\omega_- t})
     \sigma^-_je^{-i\omega_{q_j}t}e^{i\alpha_j cos(\omega_j t+\varphi)}      \nonumber\\
& &+\sum_{j=N+1}^{2N}\frac{g_j}{2\sqrt{2}}(P_+^+ e^{i\omega_+ t} -P_-^+e^{i\omega_- t})
     \sigma^-_je^{-i\omega_{q_j}t}e^{i\alpha_j cos(\omega_j t+\varphi)}      \nonumber\\
& &+\sum_{j=1}^{2N}\frac{\Omega_j}{2} e^{-i(\omega_{d_j} t+\varphi_d)}\sigma_j^+e^{i\omega_{q_j}t}e^{-i\alpha_j cos(\omega_j t+\varphi)} +H.c.
\end{eqnarray}
where $\alpha_j=A_j/\omega_j$. Note that
\begin{eqnarray}
e^{i\alpha_j cos(\omega_j t+\varphi)} \equiv
\sum\limits_{m=-\infty}^\infty i^m J_m(\alpha_j)e^{im(\omega_j t+\varphi)},
\end{eqnarray}
thus setting  $\omega_{d_j}=\omega_{q_j}$, $\varphi=\pi/2$,  and
\begin{eqnarray}\label{condition}\label{7}
|\omega_--\omega_{q_j}+m\omega_j \gg \{|\delta|, g_jJ_1(\alpha_j) \}, \quad
|\omega_+-\omega_{q_j}+m\omega_j| \gg \{|\delta|, g_jJ_1(\alpha_j) \}
\end{eqnarray}
with $\delta=\omega_--\omega_{q_j}-\omega_j$ and $(m\neq-1)$, we can neglect the terms oscillating fast, thus  Eq. (\ref{hw}) reduces to
\begin{eqnarray}\label{Heff}\label{6}
H_{w}'=\sum_{j=1}^{2N}\frac{J_0(\alpha_j)\Omega_j}{2}\sigma_j^x
+ \frac{1}{2\sqrt{2}} \left[\sum_{j=1}^{N}g_j J_1(\alpha_j)e^{i\delta t}P^{\dagger}_-\sigma^-_j
-\sum_{j=N+1}^{2N} g_j J_1(\alpha_j)e^{i\delta t}P^{\dagger}_-\sigma^-_j+H.c.\right].
\end{eqnarray}

Defining $H_{w_0}=\sum_{j=1}^{2N}\frac{J_0(\alpha_j)\Omega_j}{2}\sigma_j^x$. In the interaction picture, the interacting Hamiltonian will be \cite{effh}
\begin{eqnarray}\label{8}
H_{int}'&=& e^{i\delta t}P^{\dagger}_-\sum_{j=1}^{N}\frac{g_j}{4\sqrt{2}}J_1(\alpha_j)(\sigma_j^x
+|-\rangle_j\langle+|e^{-iJ_0(\alpha_j)\Omega_j t}-|+\rangle_j\langle-|e^{iJ_0(\alpha_j)\Omega_j t})\nonumber\\
&&-e^{i\delta t}P^{\dagger}_-\sum_{j=N+1}^{2N}\frac{g_j}{4\sqrt{2}}J_1(\alpha_j)( \sigma_j^x
 + |-\rangle_j\langle+|e^{-iJ_0(\alpha_j)\Omega_j t}- |+\rangle_j\langle-|e^{iJ_0(\alpha_j)\Omega_j t})+H.c.
\end{eqnarray}
Assuming that $J_0(\alpha_j)\Omega_j\gg \{\delta, J_1(\alpha_j)g_j\}$, and eliminate
the oscillate with high frequencies, $H_{int}'$ reduces to
\begin{eqnarray}\label{9}
H_{int}  &=& \frac{e^{i\delta t}}{4\sqrt{2}} \left[\sum_{j=1}^{N} g_j J_1(\alpha_j) P^{\dagger}_-   \sigma_j^x
-\sum_{j=N+1}^{2N} g_j J_1(\alpha_j) P^{\dagger}_-  \sigma_j^x \right] +H.c..
\end{eqnarray}
\end{widetext}
Note that, in the case, we can control of the effective coupling strength $g_j$ by varying the externally the classical field, i.e., by controlling the amplitude $\alpha_j$, so that
\begin{eqnarray}  \label{gjalpha}
g_jJ(\alpha_j)/\sqrt{2}=g
\end{eqnarray}
can be met. Then $H_{int}$ can be written in the form of
\begin{eqnarray}\label{10}
H_{int} &=&\frac{g}{2} P^{\dagger}_- J_ x e^{i\delta t} +H.c.,
\end{eqnarray}
where we have set $J_x=J_1^x-J_2^x$ with $J_1^x=\frac{1}{2}\sum_{j=1}^{N}\sigma_j^x$ and $J_2^x=\frac{1}{2}\sum_{j=N+1}^{2N}\sigma_j^x$.

The evolution operator of the effective Hamiltonian reads
\begin{eqnarray} \label{u}\label{11}
U (\gamma) = \exp\left({i\gamma J_x^{2}}\right)
\exp\left({iB P_-^{\dagger}J_x}\right)
\exp\left({i{B^{\ast}} P_-J_x}\right), \notag\\
 \end{eqnarray}
where
\begin{eqnarray}
\gamma&=&\frac{g ^{2}}{4\delta}\left[t + \frac{1}{i\delta}\left( e^{-i\delta t}-1\right)\right], \notag\\
B&=&\frac{g }{2i \delta}\left(e^{i\delta t} -1 \right).
 \end{eqnarray}
It is obvious that $B(t)$ is a periodic function of time
and vanishes at $\delta \tau=2k\pi$ where $ k=1,2,3,...$. At those time intervals, the evolution operator in Eq. (\ref{u}) reduces to
\begin{eqnarray}\label{u1}\label{12}
U(\tau)=\exp\left[i \gamma(\tau) J_x ^{2}\right],
\end{eqnarray}
which can be directly used to generate GHZ states when $\gamma(\tau) =\pi/2$. In this case, $\delta =\sqrt{k}g, \quad \tau=2\pi\sqrt{k}/g$. For a fast scheme, we can set $k=1$.

For an initial state
$|\Psi(0)\rangle=|00\cdots 0\rangle_a\bigotimes |00\cdots 0\rangle_b$ for $2N$ qubits in two cavities a and b, the final state is found to be a GHZ state of
\begin{eqnarray}\label{GHZ}\label{13}
|\Psi(\tau)\rangle_N &=& U\left(\frac{\pi}{2}\right)|\Psi(0)\rangle  \\
&=&\frac{1}{\sqrt{2}}\left[
|00\cdots0\rangle_a|00\cdots0\rangle_b
-i|11\cdots 1\rangle_a|11\cdots 1\rangle_b\right], \notag
\end{eqnarray}
when \emph{N} is even; the detail derivation are presented in  Appendix A.

\section{Numerical simulations}
In order to obtain the effective Hamiltonian, several approximations have been made here According to the postulated conditions in Eq. (\ref{7}), we need to choose a suitable value of the frequency $\omega_j$ of \emph{z}-direction magnetic. $\omega_j$ is set to around a fixed value, which should be adjusted with respect to  the corresponding qubit frequency $\omega_{qj}$, in order  to ensure $\delta$ to be equal for different qubits. For simplicity, we set $\omega_j/2\pi=2\pi\times 600$ MHz in our numerical simulation.  Meanwhile, suitable driving amplitudes $A_j$s of the \emph{z}-direction classical fields can be chosen to make sure that the effective qubit-cavity coupling strengths to be the same, as shown in Eq. (\ref{gjalpha}), in spite of the fact that the original qubit-cavity coupling strength $g_j$s are not identical. Here,   $A_j$s are used to tune $g_j$ to a same value $g/2\pi=$15 (10) MHz in the two (four) qubits case. In addition, since $\alpha_j$ becomes a certain value, we also choose the amplitudes of each \emph{x}-direction magnetic flux $\Omega_j$ to meet the condition of $J_0(\alpha_j)\Omega_j=\Omega$.
For our simulation, the used master equation is
\begin{eqnarray}
\frac{\mathrm{d}\rho(t)}{\mathrm{d}t}&=& -i\left[H_w,\rho(t)\right]+\kappa L(a)+\kappa L(b) \notag\\
&&+\sum_{j=1}^{2N} \left[ \beta L\left(\sigma_j^-\right) + \gamma L\left(\sigma_j^z\right)\right]
\end{eqnarray}
where
$L\left(B\right)=B\rho(t)B^\dag - B^\dag B\rho(t)/2-\rho(t)B^\dag B/2$  is the Lindblad operator with $B\in\{a, b, \sigma_j^z, \sigma_j^-\}$.
Here, the  decay and dephasing rates for all the qubits and the decay for both cavities are all set to be equal as $\kappa=\beta=\gamma=2\pi \times 4$ kHz.

\begin{figure}[tbp]
  \centering
  \includegraphics[width=\linewidth]{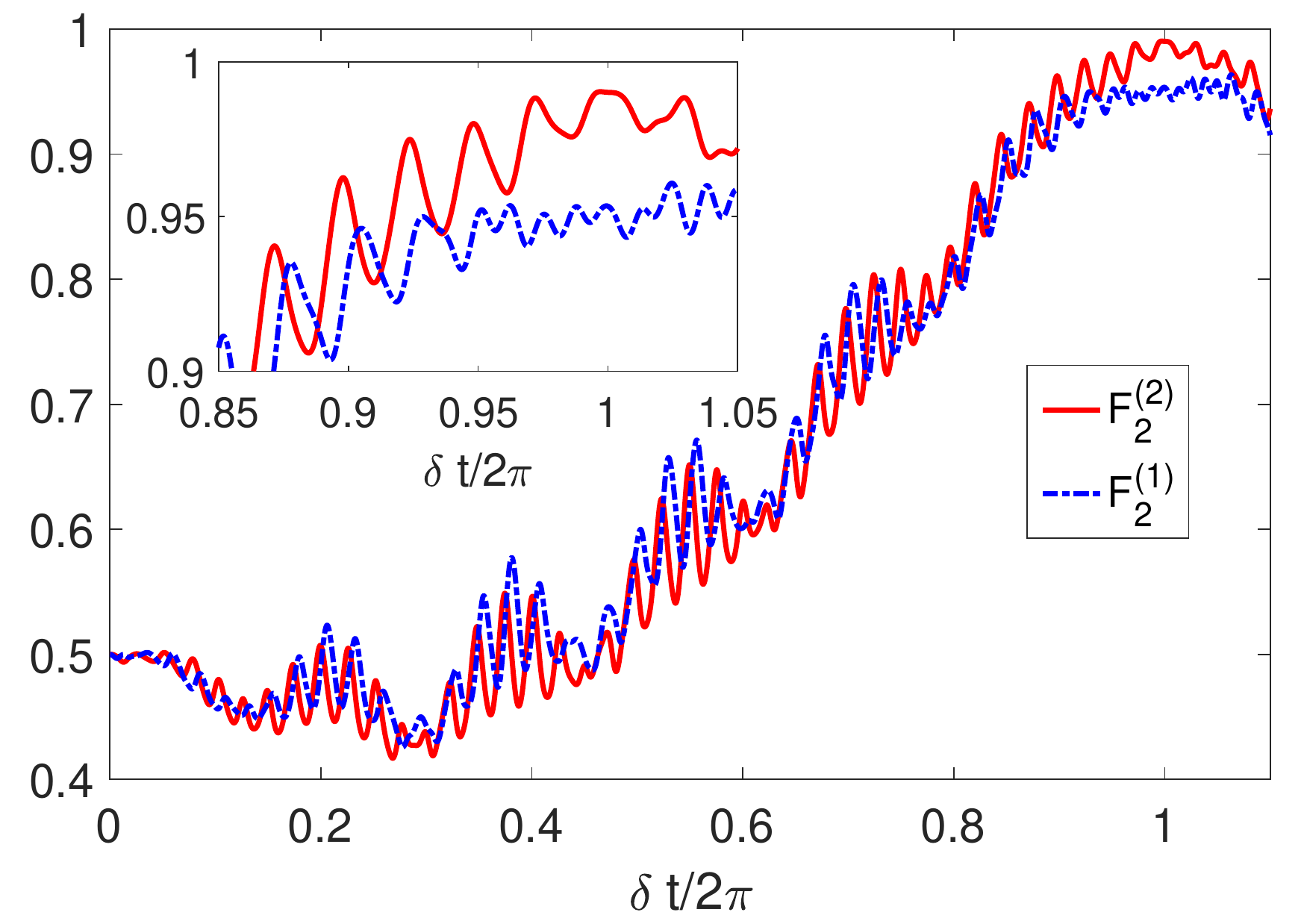}
  \caption{Time dependence of the fidelity of the two qubits entangled state generation for the two coupled cavities (red solid line) and one cavity (blue dashed line) cases.}\label{ene}
\end{figure}

\begin{table}[b]
\centering
\caption{The chosen parameters for our two qubits entangled state generation schemes.}
\begin{tabular}{|c|c|c|c|}
\hline
                  &   two-cavities case & one-cavity case  \\
\hline
         $\alpha_1$     &  0.5171 & 0.7025  \\
\hline
         $\alpha_2$     &  0.5036  & 0.6845   \\
\hline
         $g_1/2\pi$     &  84.9 MHz & 45  MHz \\
\hline
         $g_2/2\pi$     &  87.0 MHz  & 46.5  MHz  \\
\hline
    $A_1/2\pi$         & 310.26 MHz & 421.5 MHz  \\
\hline
    $A_2/2\pi$       &  302.16 MHz & 410.7 MHz  \\
\hline
    $\Omega_1/2\pi$     & 96.3 MHz  & 102.2 MHz    \\
\hline
    $\Omega_2/2\pi$     & 96.0 MHz      & 101.6 MHz   \\
\hline
    $\Omega/2\pi$       & 90.0 MHz & 90.0 MHz  \\
\hline
\end{tabular}
\end{table}

As shown in Fig. 2, the fidelities of two qubits entangled state generation are plotted with respective to time, for both the one and two cavities cases and the other parameters are list in table I.
Obviously in the Fig.2, when the $\delta\tau=2\pi$, the fidelity of red solid line reaches up to 99.02\% which is larger than the same data of the blue dashed line. This proves that the performance of generating GHZ state by a z-direction biasing   magnetic flux in the two-cavities case is better than that of the one cavity case. Note that, the quantity of inter-cavity  coupling $J$ exists only in the two-cavities case, and thus in the  one cavity group we compensate this to the qubits' frequency, so that the \emph{z}-direction driving frequency $\omega_{j}/2\pi$ is still equal to 600MHz, be consistent with that of the  two-cavity case. Besides the number of cavities, decoherence and the oscillating terms are non-negligible factors, which decrease the fidelity of the  generated GHZ state.
In order to make all qubit-cavity coupling strength $g_j$ to be a same value, each qubit is driven by the classical field along \emph{z} directions which bring in the oscillating terms that decreases the fidelity about $0.5\% $. Theoretically, with the increasing of the frequency $\omega_j$, the affect of the oscillating terms would decrease.

\begin{figure}
  \centering
  \includegraphics[width=\linewidth]{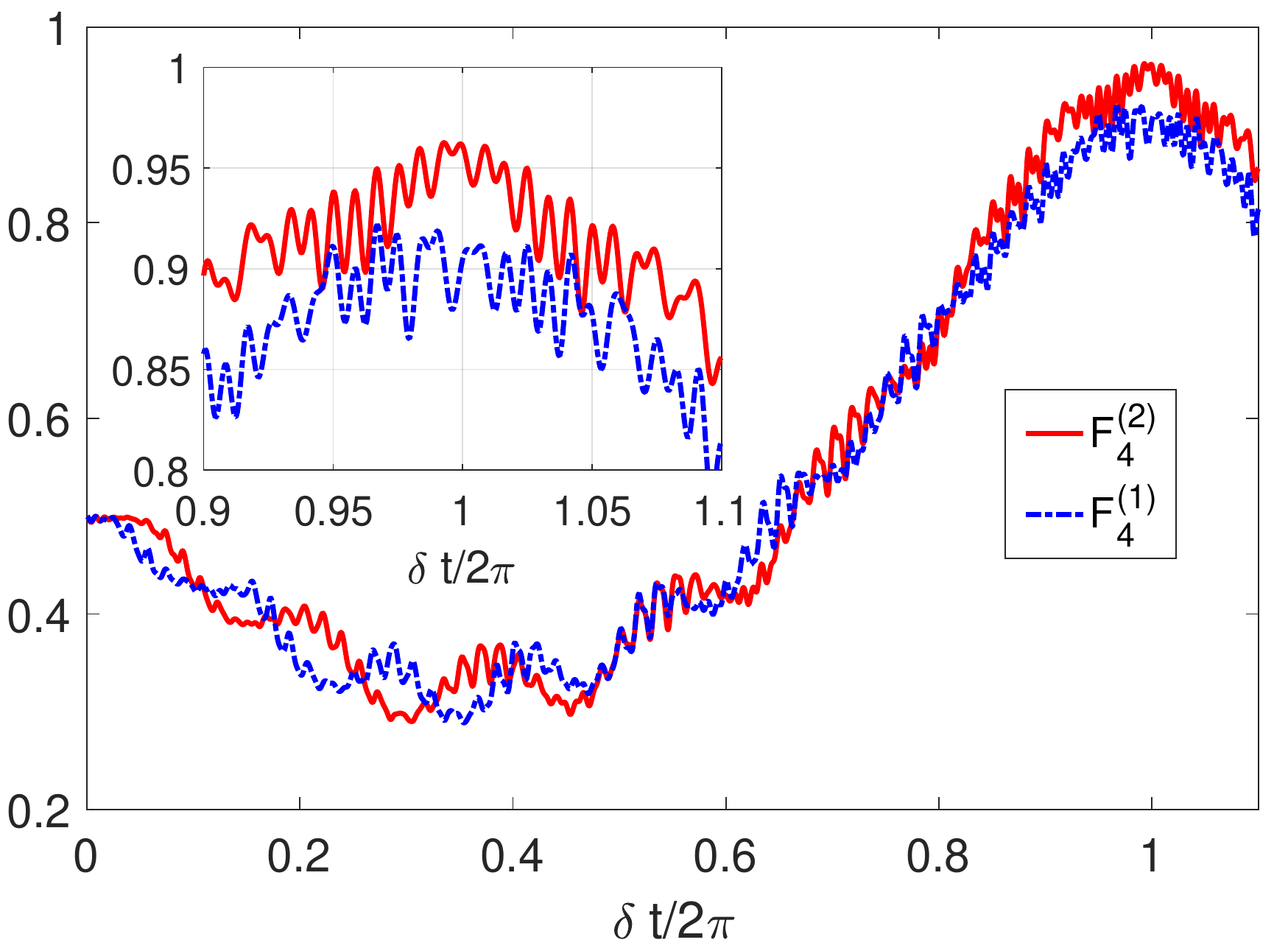}
  \caption{Time dependence of the fidelity of the four qubits GHZ state generation for the two coupled cavities (red solid line) and one cavity (blue dashed line) cases.            }\label{ene}
\end{figure}

\begin{table}[tb]
\centering
\caption{The chosen parameters for our four qubits entangled state generation shcmes.}
\begin{tabular}{|c|c|c|c|}
\hline
                  &   two-cavities case & one-cavity case  \\
\hline
         $\alpha_1$     &  1.0134 & 0.7698  \\
\hline
         $\alpha_2$     &  0.9837  & 0.7389   \\
\hline
         $\alpha_3$     &  0.9562 & 0.7025  \\
\hline
         $\alpha_4$     &  0.9303  & 0.6845   \\
\hline
         $g_1/2\pi$     &  31.8 MHz & 28.0 MHz \\
\hline
         $g_2/2\pi$     &  32.5 MHz  & 29.0 MHz  \\
\hline
         $g_3/2\pi$     &  33.2 MHz & 30.0  MHz \\
\hline
         $g_4/2\pi$     &  33.9 MHz  & 31.0  MHz \\

\hline
    $A_1/2\pi$         & 608.0 MHz & 461.9 MHz  \\
\hline
    $A_2/2\pi$       &  590.2 MHz & 443.3 MHz  \\
\hline
    $A_3/2\pi$         & 573.7 MHz & 421.5 MHz  \\
\hline
    $A_4/2\pi$       &  558.2 MHz & 410.7 MHz  \\

\hline
    $\Omega_1/2\pi$     & 79.0 MHz  & 93.3 MHz    \\
\hline
    $\Omega_2/2\pi$     & 77.7 MHz  & 92.2 MHz   \\
\hline
    $\Omega_3/2\pi$     & 76.5 MHz  & 90.9 MHz    \\
\hline
    $\Omega_4/2\pi$     & 75.5 MHz  & 90.3 MHz   \\
\hline
    $\Omega/2\pi$       & 60.0 MHz & 80.0 MHz  \\
\hline
\end{tabular}
\end{table}

As shown in Fig. 3, the fidelities of four qubits entangled state generation are plotted with respective to time, for both the one and two cavities cases and the other parameters are list in table II. When the $\delta\tau=2\pi$, the value of fidelity is 96.1\% and 90.5\% for the two- and one-cavity cases, respectively. The same conclusion we can get is that the  two-cavity case shows  better performance. Therefore, keeping all qubits apart to each cavity will beneficial to the increase of fidelity. This is because that the  cross talk effect among the oscillating terms will be separated into to cavities and thus some of them have been suppressed.

\section{CONCLUSION}
In summary, we have put forward a project to generate GHZ state in two coupled cavities scenario. In this method, each qubit is driven simultaneously by the classical fields. Obviously, the role of the \emph{x}-direction classical field is manipulating the qubit state. Therefore, we mainly adjust the qubit-cavity strength $g_j$ by controlling the amplitude and frequency of the \emph{z}-direction classical field. Because of deviation of superconducting manufacturing technique or the other environmental factor, we need a way to make all qubit-cavity coupling strength $g_j$ to be a same value, which can be achieved here by selecting the appropriate  parameters. In addition, we also show that the coupled cavities scenario is better than a single cavity case. Therefore, our proposal represents a promising alternative for multipartite entangled states generation with superconducting qubits.

\section*{Acknowledgments}
This work was supported by the National Natural Science Foundation of China (Grant No. 11874156) and the National Key R \& D Program of China (Grant No. 2016YFA0301803).

\appendix
\section{Derivation details for Eq. (\ref{GHZ})}
In this Appendix, we present some derivation details in the maintext.   We note that
\begin{eqnarray}
&&|\Psi(\tau)\rangle_N=e^{\left(i {\pi \over 2} J_ x ^{2}\right)}
\left|{N \over 2}, -{N \over 2}\right\rangle_{z,a}
\left|{N \over 2}, -{N \over 2}\right\rangle_{z,b}\notag\\
&=& e^{\left(i {\pi \over 2} J_ x ^{2}\right)}
\sum_{M_1, M_2} C_{M_1}\left|{N \over 2}, M_1\right\rangle_{x,a}
C_{M_2} \left|{N \over 2}, M_2\right\rangle_{x,b}\\
&=& \sum_{M_1, M_2} e^{i {\pi \over 2} \left(M_1-M_2\right)^{2} }
 C_{M_1}\left|{N \over 2}, M_1\right\rangle_{x,a}
C_{M_2} \left|{N \over 2}, M_2\right\rangle_{x,b}.\notag
\end{eqnarray}

\vspace{.2cm}
As
\begin{eqnarray}
&& \exp\left[i {\pi \over 2} \left(M_1-M_2\right)^{2}\right]
 = \left\{
     \begin{array}{ll}
       i, &  M_1-M_2 \quad is \quad odd; \\
       1, & M_1-M_2 \quad is \quad even.
     \end{array}
     \right.\notag\\
&=&\frac{1}{\sqrt{2}} \left[e^{i{\pi \over 4}}
+(-1)^{\left(M_1-M_2\right)}e^{-i{\pi \over 4}}\right]\notag\\
&=&\frac{1}{\sqrt{2}} \left[e^{i{\pi \over 4}}
+(-1)^{2N-2M_1+\left(M_1-M_2\right)}e^{-i{\pi \over 4}}\right]\\
&=&\frac{1}{\sqrt{2}} \left[e^{i{\pi \over 4}}
+(-1)^N (-1)^{(N/2-M_1)} (-1)^{(N/2-M_2)}e^{-i{\pi \over 4}}\right],\notag
\end{eqnarray}
we get
\begin{widetext}
\begin{eqnarray}
|\Psi(\tau)\rangle_N
&=& \frac{1}{\sqrt{2}} \sum_{M_1, M_2}
\left[e^{i{\pi \over 4}}
+(-1)^N (-1)^{(N/2-M_1)} (-1)^{(N/2-M_2)}e^{-i{\pi \over 4}}\right]
C_{M_1}\left|{N \over 2}, M_1\right\rangle_{x,a}
C_{M_2} \left|{N \over 2}, M_2\right\rangle_{x,b} \notag\\
&=&\frac{1}{\sqrt{2}} \left[e^{i{\pi \over 4}}
\sum_{M_1} C_{M_1}\left|{N \over 2}, M_1\right\rangle_{x,a}
\sum_{M_2} C_{M_2} \left|{N \over 2}, M_2\right\rangle_{x,b}\right.\notag\\
&&+\left. (-1)^N e^{-i{\pi \over 4}}
\left(\sum_{M_1} (-1)^{(N/2-M_1)} C_{M_1}
\left|{N \over 2}, M_1\right\rangle_{x,a}\right)
\left(\sum_{M_2} (-1)^{(N/2-M_2)} C_{M_2}
\left|{N \over 2}, M_2\right\rangle_{x,b}\right)\right] \notag\\
&=&\frac{1}{\sqrt{2}}\left[e^{i\frac{\pi}{4}}
\left|{N \over 2}, -{N \over 2}\right\rangle_{z,a}
\left|{N \over 2}, -{N \over 2}\right\rangle_{z,b}
+(-1)^N e^{-i\frac{\pi}{4}}
\left|{N \over 2}, {N \over 2}\right\rangle_{z,a}
\left|{N \over 2}, {N \over 2}\right\rangle_{z,b}\right],\notag\\
&=&\frac{1}{\sqrt{2}}\left[
e^{i\frac{\pi}{4}}|00\cdots0\rangle_a|00\cdots0\rangle_b
+(-1)^N e^{-i\frac{\pi}{4}}|11\cdots 1\rangle_b|11\cdots 1\rangle_b\right].
\end{eqnarray}

When $N=1$, the final state is
\begin{eqnarray}
|\Psi(\tau)\rangle_1 = \frac{1}{\sqrt{2}}\left(|0\rangle_a|0\rangle_b
+i|1\rangle_a|1\rangle_b\right).
\end{eqnarray}
In the Schr\"{o}dinger picture, the above state can be written as
\begin{eqnarray}
&\frac{1}{\sqrt{2}}\left[
(e^{-i\frac{\Omega}{2}t}|+\rangle_a+e^{i\frac{\Omega}{2}t}|-\rangle_a)
(e^{-i\frac{\Omega}{2}t}|+\rangle_b+e^{i\frac{\Omega}{2}t}|-\rangle_b)
+i(e^{-i\frac{\Omega}{2}t}|+\rangle_a-e^{i\frac{\Omega}{2}t}|-\rangle_a)
(e^{-i\frac{\Omega}{2}t}|+\rangle_b-e^{i\frac{\Omega}{2}t}|-\rangle_b)\right]
\end{eqnarray}
where
$|\pm\rangle_a= e^{i\Theta_1} |0\rangle \pm  e^{-i\Theta_1}|1\rangle$ and
$|\pm\rangle_b= e^{i\Theta_2} |0\rangle \pm  e^{-i\Theta_2}|1\rangle$
with $\Theta_n= \omega_{q_n} t/2 - \alpha_n \cos(\omega_n t+\varphi)/2$ and $n\in\{1, 2\}$.

\end{widetext}

\end{document}